\documentclass[conference]{IEEEtran}
\IEEEoverridecommandlockouts

\usepackage{cite}
\usepackage{amsmath,amssymb,amsfonts}
\usepackage{algorithmic}
\usepackage{graphicx}
\usepackage{textcomp}
\usepackage[dvipsnames]{xcolor}

\usepackage{float}
\usepackage[hyphens]{url}
\usepackage{enumitem}
\usepackage[final]{microtype}
\usepackage{subfig}
\usepackage{pifont}
\usepackage{listings}
\usepackage{hyperref}
\usepackage{cleveref}

\definecolor{codegreen}{rgb}{0,0.6,0}
\definecolor{codegray}{rgb}{0.5,0.5,0.5}
\definecolor{codepurple}{rgb}{0.58,0,0.82}
\lstdefinestyle{long_code}{
    commentstyle=\color{codegreen},
    keywordstyle=\color{magenta},
    numberstyle=\tiny\color{codegray},
    stringstyle=\color{codepurple},
    basicstyle=\ttfamily\footnotesize,
    breakatwhitespace=false,
    breaklines=true,
    captionpos=b,
    keepspaces=true,
    numbers=left,
    numbersep=5pt,
    showspaces=false,
    showstringspaces=false,
    showtabs=false,
    tabsize=2,
    frame=single,
}
\lstdefinestyle{short_code}{
    basicstyle=\ttfamily\small,
}

\def\BibTeX{{\rm B\kern-.05em{\sc i\kern-.025em b}\kern-.08em
    T\kern-.1667em\lower.7ex\hbox{E}\kern-.125emX}}
\begin{document}

\title{Timing the Transient Execution:\\ A New Side-Channel Attack on Intel CPUs % via EFLAGS register and Jcc Instructions
}

\author{
\IEEEauthorblockN{
    Yu Jin\IEEEauthorrefmark{1},
    Pengfei Qiu\IEEEauthorrefmark{1},
    Chunlu Wang\IEEEauthorrefmark{2},
    Yihao Yang\IEEEauthorrefmark{3}, \\
    Dongsheng Wang\IEEEauthorrefmark{4},
    Gang Qu\IEEEauthorrefmark{5}} \\

    \IEEEauthorblockA{
        \IEEEauthorrefmark{1}\IEEEauthorrefmark{2}\IEEEauthorrefmark{3}Key Laboratory of Trustworthy Distributed Computing and Service (BUPT), Ministry of Education~~ \\
        \IEEEauthorrefmark{4}Tsinghua University~~
        \IEEEauthorrefmark{5}University of Maryland~~
    }
    \IEEEauthorblockA{lambda.jinyu@gmail.com, \{qpf, wangcl\}@bupt.edu.cn, khaosyg@gmail.com,}
    \IEEEauthorblockA{wds@tsinghua.edu.cn, gangqu@umd.edu}
}

\maketitle

\begingroup\renewcommand\thefootnote{*}
\begin{NoHyper}
\footnotetext{Equal contributions.}
\end{NoHyper}
\endgroup

\pagestyle{plain}
\pagestyle{empty}

\begin{abstract}
    The transient execution attack is a type of attack leveraging the vulnerability of modern CPU optimization technologies. New attacks surface rapidly. The side-channel is a key part of transient execution attacks to leak data.
    % EFLAGS register in Intel CPUs includes various flags for status representation, control, and other purposes. In condition instructions such as Jcc (jump on condition code), the execution relies on Status Flags like ZF.

    In this work, we discover a vulnerability that the change of the EFLAGS register in transient execution may have a side effect on the Jcc (jump on condition code) instruction after it in Intel CPUs. Based on our discovery, we propose a new side-channel attack that leverages the timing of both transient execution and Jcc instructions to deliver data. This attack encodes secret data to the change of register which makes the execution time of context slightly slower, which can be measured by the attacker to decode data. This attack doesn't rely on the cache system and doesn't need to reset the EFLAGS register manually to its initial state before the attack, which may make it more difficult to detect or mitigate. We implemented this side-channel on machines with Intel Core i7-6700, i7-7700, and i9-10980XE CPUs. In the first two processors, we combined it as the side-channel of the Meltdown attack, which could achieve 100\% success leaking rate. We evaluate and discuss potential defenses against the attack. Our contributions include discovering security vulnerabilities in the implementation of Jcc instructions and EFLAGS register and proposing a new side-channel attack that does not rely on the cache system.
\end{abstract}


\section{Introduction}
The increasing complexity and aggressive optimizations of modern CPUs, with their many microarchitectural features, have led to improved performance, but they have also created a range of security vulnerabilities\cite{holtryd2022sok,Intel}. This complexity and optimization are the root cause of many security issues, including side-channel attacks\cite{liu2015last}, Meltdown attack\cite{Lipp2018,Bulck2018}, Spectre attack\cite{Kocher2018,Koruyeh2018}, Microarchitectural data sampling (MDS) attack \cite{RIDL, Fallout, ZombieLoad}, fault injection attack\cite{plundervolt,clkscrew,qiu2019voltjockey,voltjockey-TZ,voltjockey-SGX, voltpwn}, and more. The complex and dynamic nature of modern CPUs has made them a challenging target for security researchers and developers to discover and mitigate, and a constant source of concern for users. As the field of computer security continues to evolve, new techniques and countermeasures will be needed to keep pace with the ever-evolving threat landscape.

\begin{figure*}[!t] 
    \centering
    \begin{minipage}[b]{0.82\textwidth}
    \centering
    \includegraphics[width=1\columnwidth]{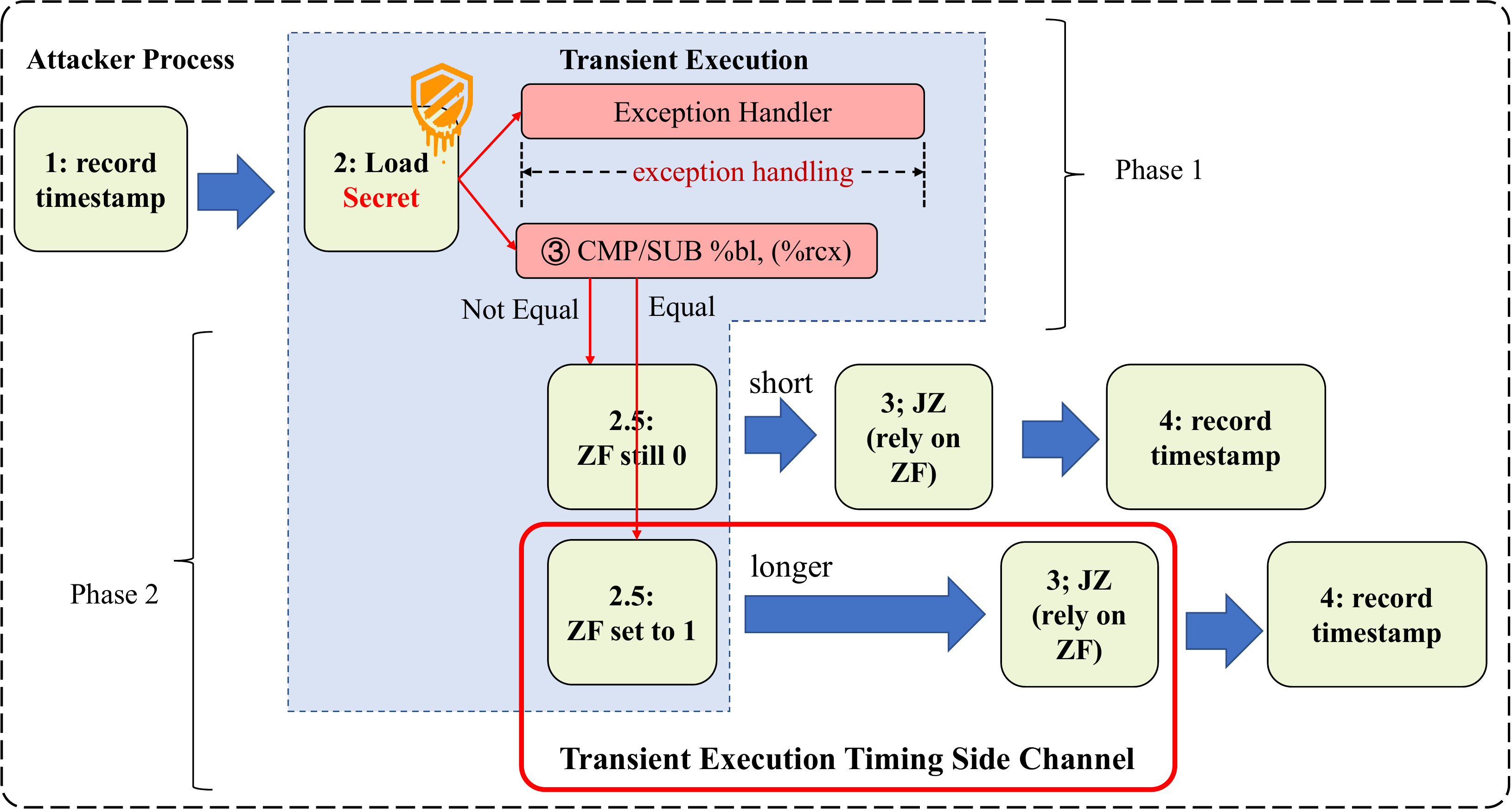}
    \end{minipage} 
    \caption{Overview of Transient Execution Timing Side-Channel.} 
    \label{fig:attack_overview}
\end{figure*}

A number of the microarchitectural side-channel attacks are based on the side effects and state of the cache system. There are many ways to leak info through the cache system. After the first cache-timing attack reported by Bernstein in 2005\cite{bernsteincache}. Multiple variations such as evict+time (2006)\cite{Osvik2006}, flush+reload(2014)\cite{Yarom14}, prime+probe(2015)\cite{liu2015last}, flush+flush(2016)\cite{gruss2016flush+}. New cache side-channel attacks are still being discovered, e.g., attacks in cache replacement policies(2020)\cite{briongos2020reload+}, streamline(2021)\cite{saileshwar2021streamline}, using cache dirty states(2022)\cite{cui2022abusing}, cache coherence(2022)\cite{guo2022adversarial}. Besides, some attacks do not directly rely on the cache system, such as PortSmash\cite{aldaya2019port_portsmash}, PlatyPus(2021)\cite{lipp2021platypus}, PMU-Spill\cite{qiu2022pmu}.

Transient execution attacks\cite{canella2019sokattacks}, including Meltdown, Spectre, and MDS attacks, exploit the complex and aggressive optimizations\cite{ZombieLoad} of modern CPUs to leak sensitive information through transient states. As a high-level overview\cite{canella2019sokattacks}, transient execution attacks consist of five phases: (1) microarchitectural preparation, (2) triggering a fault, (3) encoding secret data to a covert channel, (4) flushing transient instructions, and (5) decoding the secret data. The success of phases 3 and 4 depends on side-channels, which require the channel state to be initialized or set to a specific state. For example, in the flush+reload attack\cite{Yarom14}, the attacker needs to flush the monitored memory line from the cache in phase 1 and encode secret data by loading one index of the memory line into the cache in phase 4. This allows the attacker to measure the time of the monitored memory line being loaded in the cache in phase 5 to decode data.

Reverse engineering efforts, such as those by uops.info\cite{abel2019uops}, have attempted to reveal information about the behavior of a processor's microarchitecture despite the lack of publicly available implementation details. Intel's manual\cite{intelsys} has provided additional insights into instruction performance characteristics, highlighting that certain instructions can cause pipeline stalls or other effects due to their functional requirements\cite{Degenbaev2012}. For example, the \texttt{MFENCE} instruction introduces a stall in the pipeline until all previous memory operations have been completed. Awareness of these nuances is crucial for optimizing code to maximize performance on a particular processor architecture and identify potential security vulnerabilities.

In our work, we conduct in-depth research on the behavior and side effects of transient execution attacks and discover a vulnerability of the implementation in Intel CPUs. Specifically, the change of the EFLAGS register in transient execution may influence the Jcc instruction after it. Based on our discovery, we introduce a novel side-channel attack that leverages the timing of transient execution with Jcc instruction. The change of EFLAGS in transient execution could make some Jcc instructions after it slightly slower. As we showed in Fig.\ref{fig:attack_overview}, by encoding secret data to the EFLAGS register, we can measure the execution time of the Jcc instruction's context to decode data without the need to reset the EFLAGS register to its initial state in phase 1 of transient attack.

This attack does not rely on the cache system, which may make it more difficult to detect compared to previous side-channel attacks. We implement this side-channel in real machines with Intel Core i7-6700 and i7-7700 and i9-10980XE CPUs. We build the Meltdown attack with our side-channel attack and evaluate it on i7-6700 and i7-7700. In practice, our side-channel can achieve 100\% success rate.

To mitigate this attack, we propose several practical mitigation methods based on our evaluation.

Our research makes several contributions:
\begin{enumerate}
    \item We discover security vulnerabilities in the implementation of EFLAGS register and Jcc Instruction. The change of EFLAGS register during transient execution can affect the timing of Jcc instruction after it.
    \item We propose a novel side-channel attack that exploits the timing of execution affected by Jcc instructions, which are dependent on the EFLAGS register. Our attack is distinct from previous side-channel attacks appearing in that it does not rely on the cache system and does not require resetting the initial state during the preparation phase of the transient execution attack.
    \item We implement this side-channel in Intel Core i7-6700 and i7-7700 and i9-10980XE CPUs. We build the Meltdown attack with this side-channel on Intel Core i7-6700 and i7-7700 CPUs on a real machine and it could achieve 100\% success rate.
\end{enumerate}

To the best of our knowledge, this is the first time that the EFLAGS register has been used as a side-channel. We hope that our work can help to improve the security of future CPUs and bring insight for microarchitecture attack research. The source code of our attacks would be published at {\url{https://github.com/}}.

The rest of the paper is organized as follows. In Section \ref{sec:background}, we introduce the background knowledge of side-channel attacks and transient execution attacks. In Section \ref{sec:attack}, we present the details of the attack. In Section \ref{sec:evaluation}, we evaluate our attack on Intel CPU. In Section \ref{sec:mitigation}, we propose several mitigations for this side-channel. In Section \ref{sec:discussion}, we discuss the limitation of this attack and future work. Finally, we conclude our work in Section \ref{sec:conclusion}.

\section{Background}
\label{sec:background}

\subsection{Microarchitecture}
The microarchitecture\cite{WikiChip} of each core of the CPU is composed of several components, such as the cache system, the frontend that includes the branch predictor, the out-of-order execution unit, etc. The CPU core is the core of the CPU, which is responsible for the execution of the instructions. The cache system is used to store the data and instructions that are frequently accessed. The frontend is responsible for the instruction fetch and decode. The out-of-order execution unit is responsible for the out-of-order execution of the instructions. As the modern CPU is an microarchitecture complex system. For most commercial CPUs, the microarchitecture of the CPU is a black box, and much research trying to reverse engineering in it\cite{Gras2020,Weber2021,Lustig2014}.

\subsection{Side-Channel Attacks}
Side-Channel Attacks in microarchitecture\cite{Rodrigo2021} is a class of attacks that exploit the side effects of a program to leak information about the program's execution. The side effects can be the cache system\cite{bernsteincache,Osvik2006,Yarom14, liu2015last, gruss2016flush+, briongos2020reload+,gras2018translation}, the branch predictor\cite{branchscope, coronado2021branchboozle}, the power\cite{lipp2021platypus}, etc. For example, the cache system can be used to leak info about the memory access pattern of the program. The branch predictor can be used to leak info about the control flow of the program. The root cause\cite{holtryd2022sok} of most of the side-channel attacks in microarchitecture is the shared resource, which is one key to the performance optimization of the CPU.

\subsection{Transient Execution Attacks}

\begin{figure}[ht]
    \centering
    \begin{minipage}[b]{1\columnwidth}
        \includegraphics[width=1\columnwidth]{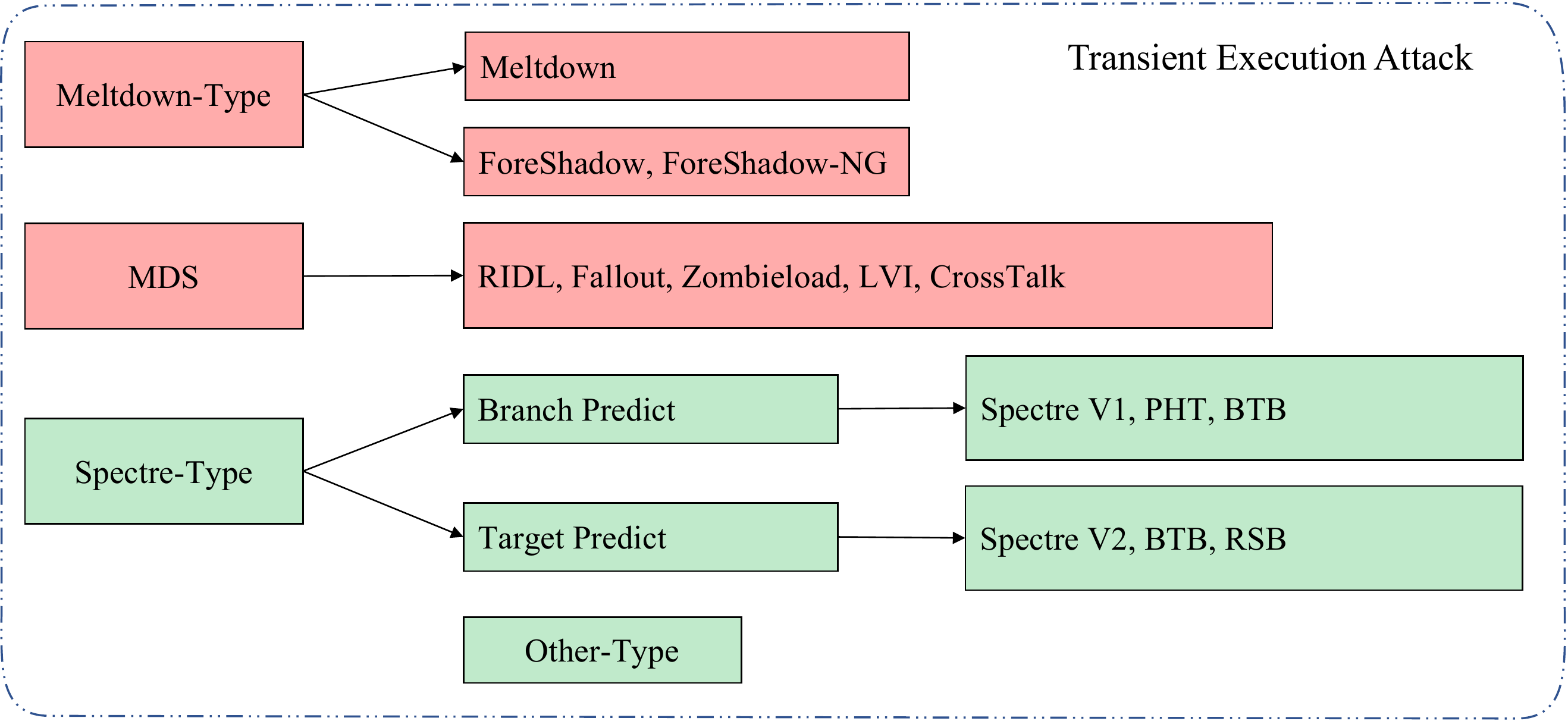}
    \end{minipage} 
    \caption{Category of Transient Execution Attacks}
    \label{fig:te-attack}
\end{figure}

Since Meltdown\cite{Lipp2018} and Spectre\cite{Kocher2018} attacks have been discovered, the transient execution attacks\cite{canella2019sokattacks} has been a hot topic in the security community. The transient execution can be caused by the fault, the branch misprediction, the cache miss, etc. And there are several variations of transient execution attacks\cite{Bulck2018, RIDL, Fallout, VanBulck2020, ZombieLoad,Ragab2021, Koruyeh2018, Maisuradze2018, Google2018, Kiriansky2018}. IP vendors like Intel and AMD have released the microcode update to mitigate the transient execution attacks\cite{IntelVulnList}. The researcher also proposed some countermeasures to mitigate it\cite{cauligi2021sok,canella2019sokattacks,holtryd2022sok,kadir2019retpoline,gonzalez2019replicating}.

\section{Transient Execution Timing Attacks}
\label{sec:attack}

\subsection{Assumption and Threat Model}

\paragraph*{Assumption} The secret data can be accessed through a transient execution attack. There are lots of transient execution attacks such as Meltdown\cite{Lipp2018}, Spectre\cite{Kocher2018}, Foreshadow\cite{Bulck2018}, ZombieLoad\cite{ZombieLoad}, etc. Although most existing attacks have been mitigated, they may have undisclosed transient execution vulnerabilities in the CPUs and have been exploited in the future\cite{qiu2022pmu}.

\paragraph*{Threat Model} The attacker runs in the unprivileged mode. And there is another victim process that runs on the same machine. We define the threat model of this study as: the attacker utilizes transient execution timing to recover the secret data acquired in the transient execution attacks.

\subsection{Attack Overview}

We implement the attack as the side-channel of the Meltdown attack, shown in Fig.~\ref{fig:attack_overview}. The attack is composed of two phases. In the first phase, we trigger transient execution and encode the secret data through the EFLAGS register. In the second phase, we measure the execution time of the Jcc instruction's context to decode data. To encode a secret through a binary flag, we need to use iteration \texttt{test\_num} to set the flag. If the \texttt{test\_num} equals the secret, the flag will be set and the secret would be encoded successfully.

\subsection{Implement Detail}

We notate the \texttt{secret\_addr} as the address of the secret data. And the \texttt{offset} is the offset of the \texttt{secret\_addr}. The \texttt{EFLAGS instruction} is the instruction that can change the EFLAGS register. The \texttt{Jcc instruction} is the instruction that can be influenced by the EFLAGS register. The available instruction set is the list in Table~\ref{tab:ins}. We use \texttt{\_\_rdtsc} from \texttt{x86intrin.h} to get the time-stamp counter of the CPU.

\begin{figure}[ht]
    \begin{center}
    \begin{minipage}{0.92\columnwidth}
    
    \begin{lstlisting}[language=c]
for (uint8_t test_num = 0; test_num <= TO; test_num++){
    start_time = __rdtsc();
    // timing context start
    asm volatile(
        "MOV %0, %%RCX;"
        "MOV %1, %%BL;"
        :
        : "r"(secret_addr + offset),
            "r"(test_num)
        :);
    if (xbegin() == (~0u))
    {
        // EFLAGS instruction
        asm volatile("SUB %BL, (%RCX);");
    }
    asm volatile(
        "JZ equal;" // Jcc instruction
        "JMP notequal;"
        "equal: NOP;"
        "notequal: NOP;");
    // timing context end
    spend_time = __rdtsc() - start_time;
    if (max_time < spend_time)
    {
        max_time = spend_time;
        argmax = i;
    }
}\end{lstlisting}
    \captionof{lstlisting}{Pseudocode for timing the transient execution attack in Intel X86 architecture.}
    \label{fig:code}
    \end{minipage}
    \end{center}
\end{figure}

\begin{figure*}[ht]
    \centering
    \begin{minipage}[b]{1\textwidth}
        \includegraphics[width=1\columnwidth]{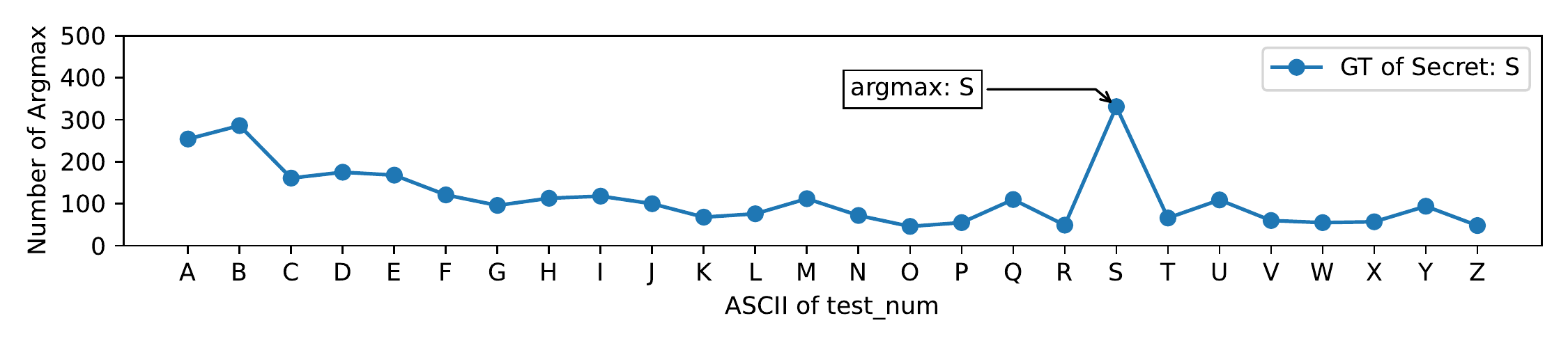}
    \end{minipage} 
    \caption{Distribution of argmax.}
    \label{fig:hist}
\end{figure*}

The \texttt{secret\_addr} is an address in kernel space. And the attacker running in an unprivileged mode which can not access the \texttt{secret\_addr}. The \texttt{offset} is the offset of the \texttt{secret\_addr}. In the TSX transaction, the attacker will try to access the \texttt{secret\_addr} by \texttt{sub} instruction. A transient execution will be caused by the fault. During the transient execution, the \texttt{ZF} may be set to 1 if the secret data in \texttt{*(secret\_addr+offest)} is equal to \texttt{i}. And the \texttt{ZF} will be restored to 0 after the transient execution. The \texttt{ZF} will be used by the \texttt{JZ} instruction to determine whether to jump to the \texttt{equal} label or the \texttt{notequal} label. In our experiment, if the secret data in \texttt{*(secret\_addr+offest)} is equal to \texttt{i}, the execution time of the context will be slower than the execution time of the context that the secret data in \texttt{*(secret\_addr+offest)} is not equal to \texttt{i}. The \texttt{max\_time} is the maximum execution time of the context. The \texttt{argmax} is the secret data, as we have shown the distribution in Fig.~\ref{fig:hist}. The \texttt{max\_time} and \texttt{argmax} can be used to decode the secret data.

\begin{table}[htbp]
    \begin{center} 
        \caption{Instruction for our Attack.}
        \label{tab:ins}
        \setlength\arrayrulewidth{1.0pt}
        \renewcommand{\arraystretch}{1.2}
        \begin{tabular}{lrrr}
            \hline
             Type & Instruction & Description & EFLAGS \\ \hline
             EFLAGS & SUB & Subtract. & ZF \\
             ~ & CMP & Compare Two Operands. & ZF \\
             ~ & CMPXCHG &  Compare and Exchange. & ZF  \\ \hline 
             Jcc & JE & Jump short if equal (ZF = 1). & ZF  \\ 
             ~ & JZ & Jump short if zero (ZF = 1). & ZF \\
             \hline
        \end{tabular}
    \end{center}  
\end{table}

\section{Experiment and Evaluation}
\label{sec:evaluation}

\subsection{Experimental Setup and Result}
We implement the attack in Intel i7-6700, i7-7700, and i9-10980XE CPU. The experiment is running in the Ubuntu 16.04 xenial with kernel version 4.15.0 (i7-6700, i7-7700) and Ubuntu 22.04 jammy with kernel version 5.15.0 (i9-10980XE). We build the Meltdown attack with our side-channel to read kernel memory from user space and achieve 100\% success leaking rate in the first two processors.

\subsection{Evaluation}

In our experiment, we found that the influence of the EFLAGS register on the execution time of Jcc instruction is not as persistent as the cache state. For about 6-9 cycles after the transient execute, the Jcc execute time will not be about to construct a side-channel.

Empirically, the attack needs to repeat thousands of times for higher accuracy. For the Listing\ref{fig:code}, we use Intel TSX, a transactional memory implementation, to completely suppress the exception. We also use system interrupt handlers to suppress the exception and achieve the same effect. The TSX is more efficient than the system interrupt handlers for transient execution attacks.

\begin{figure}[ht]
    \centering
    \begin{minipage}[b]{1\columnwidth}
        \includegraphics[width=1\columnwidth]{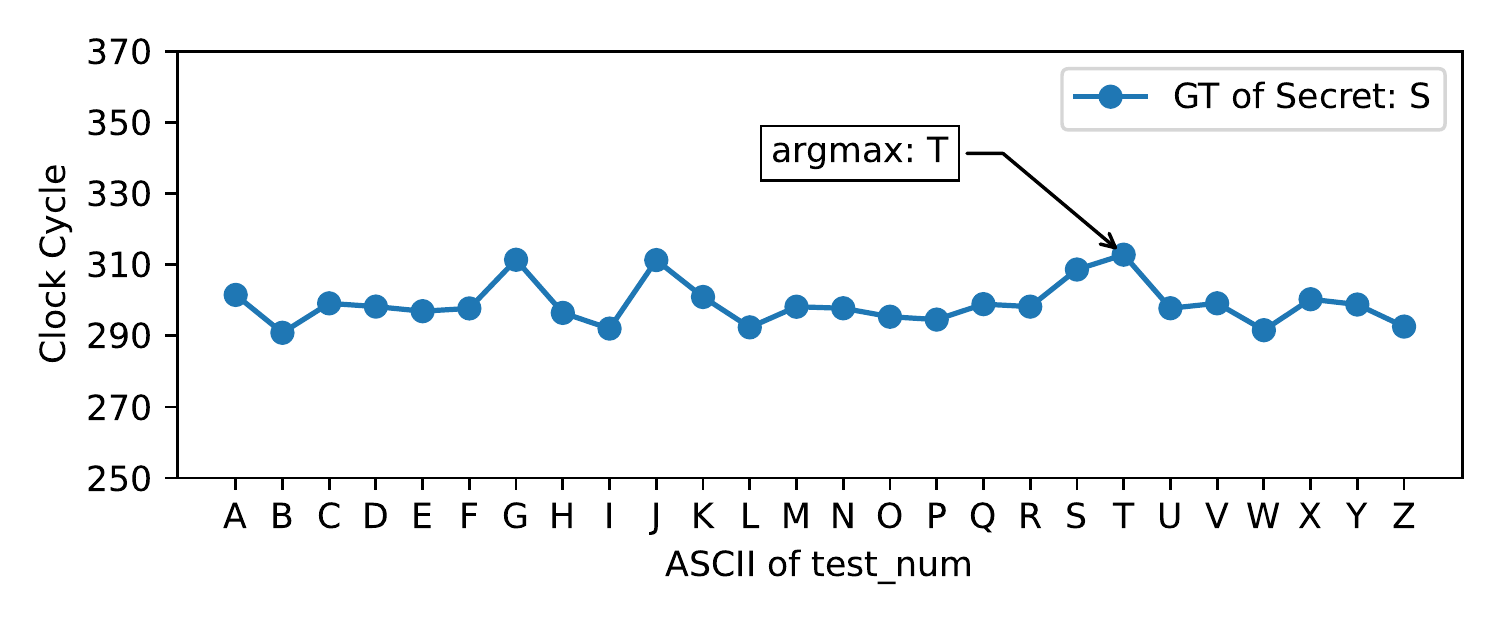}
    \end{minipage} 
    \caption{Distribution of average clock.} 
    \label{fig:mean}
\end{figure}

Though the statistical distribution of the argmax of timing can be used to decode the secret data. The distribution of the average clock, due to the noise, cannot be used as a side-channel, as we have shown in Fig.~\ref{fig:mean}.

\section{Mitigation}
\label{sec:mitigation}

For mitigating the Transient Execution Timing attacks, we can use two gadgets to delay the Jcc instruction or rewrite the EFLAGS register after the transient execution.

\subsection{Hareware Mitigation} The implementation of the Jcc instruction should not have timing or other side effects on different conditions to avoid the adversarial execution measuring.

\subsection{Delay Jcc}
If the Jcc instruction is not executed immediately after the EFLAGS register has been changed, the influence of the EFLAGS register can be reduced. 10 cycles are enough to reduce the influence of the EFLAGS register. \texttt{.rept count} Repeat the sequence of lines between the \texttt{.rept} directive and the next \texttt{.endr} directive count times. By using the \texttt{NOP} instruction to delay the Jcc instruction, we can reduce the influence of the EFLAGS register on the execution time. There are lots of ways to delay, We just give one example here.

\subsection{Rewrite EFLAGS}
The \texttt{LAHF} and \texttt{SAHF} instructions are x86 assembly language instructions that are used to manipulate the low 8 bits of the FLAGS register in the x86 processor\cite{intelsys}. The \texttt{LAHF} instruction is short for "Load AH from Flags". It loads the low 8 bits of the FLAGS register into the AH register while leaving the upper 8 bits of the AH register unchanged. The AH register is a 16-bit register that is used to store the high byte of the AX register. The \texttt{SAHF} instruction is short for "Store AH into Flags". It stores the low 8 bits of the AH register into the low 8 bits of the FLAGS register while leaving the upper 8 bits of the FLAGS register unchanged.

\begin{figure}[ht]
\begin{center}
    \begin{minipage}{0.92\columnwidth}
    \begin{lstlisting}[style=short_code, language=c]
asm volatile(
    ".rept 6 \n\t"
    "NOP \n\t"
    ".endr \n\t");\end{lstlisting}
    \captionof{lstlisting}{Pseudocode 1 for mitigating Transient Execution Timing attacks in Intel X86 architecture.}
    \label{fig:mitigation-delay}
    \end{minipage}
\end{center}
\end{figure}

The \texttt{PUSHF} and \texttt{POPF} instructions are x86 assembly language instructions that are used to push and pop the contents of the FLAGS register onto and off of the stack, respectively. The \texttt{PUSHF} instruction pushes the entire 16-bit FLAGS register onto the top of the stack. This includes the status flags that are used to indicate the result of arithmetic and logic operations, as well as other control flags that control the behavior of the processor. The \texttt{POPF} instruction pops the contents of the top of the stack into the FLAGS register. This can be used to restore the state of the FLAGS register after it has been saved by a previous \texttt{PUSHF} instruction.

By rewriting the EFLAGS though \texttt{LAHF} and \texttt{SAHF}, or \texttt{PUSHF} and \texttt{POPF} instructions, the influence of the EFLAGS register can be reduced.

\begin{figure}[ht]
\begin{center}
    \begin{minipage}{0.92\columnwidth}
    \begin{lstlisting}[style=short_code, language=c]
asm volatile(
    "LAHF;"
    "SAHF;"
    // or
    "PUSHF;"
    "POPF;" );\end{lstlisting}
    \captionof{lstlisting}{Pseudocode 2 for mitigating Adversarial Jcc attacks in Intel X86 architecture.}
    \label{fig:mitigation-rewrite}
    \end{minipage}
\end{center}
\end{figure}

\section{Discussion and Futrue Work}
\label{sec:discussion}
The root causes of this attack are still not fully understood. We guess that there is some buffer in the execution unit of the Intel CPU which need some time to revert if the execution should be withdrawn. This withdrawal process will cause a stall if the following instruction depends on the target of the buffer.

\subsection{Limitation}
This timing attack relies on other transient execution attacks to build a real-world attack and it is easy to be disturbed by noise. But it is still a new side-channel attack and worth further exploration. This attack may bring insight for new microarchitecture attacks and give a new way to build side-channel attacks in cache side-channel resistant CPU.

\subsection{Futrue Work}
\paragraph{Combined with other Transient Execution Attacks} More experiments are underway to fully understand it. Which will be published in the future.

\paragraph{Finding other various} There may be other microarchitectural components and instructions that can be used as timing side-channels. We will continue to explore this area.

\section{Conclusions}
\label{sec:conclusion}
We present a new side-channel attack that leaks info through the timing of execution. When the ZF has been changed from 0 to 1 during transient execution caused by a fault in the Meltdown attack, though the ZF will be restored to 0 after the transient execution, the instruction like JZ execution time will be slightly longer. As a result, we can leak info throw EFLAGS register by measuring the execution time of context. Compared with previous side-channel attacks, our attack does not rely on the cache system, which may make it difficult to be detected by existing tools or methods\cite{Li2022,Wang2020,Guarnieri2018}.

As far as we know, our work is the first that build side-channel with the EFLAGS register. Hope our work can help to improve the security of future CPUs.

\bibliographystyle{IEEEtran.bst}
\bibliography{refs}

\begin{thebibliography}{10}
\providecommand{\url}[1]{#1}
\csname url@samestyle\endcsname
\providecommand{\newblock}{\relax}
\providecommand{\bibinfo}[2]{#2}
\providecommand{\BIBentrySTDinterwordspacing}{\spaceskip=0pt\relax}
\providecommand{\BIBentryALTinterwordstretchfactor}{4}
\providecommand{\BIBentryALTinterwordspacing}{\spaceskip=\fontdimen2\font plus
\BIBentryALTinterwordstretchfactor\fontdimen3\font minus \fontdimen4\font\relax}
\providecommand{\BIBforeignlanguage}[2]{{%
\expandafter\ifx\csname l@#1\endcsname\relax
\typeout{** WARNING: IEEEtran.bst: No hyphenation pattern has been}%
\typeout{** loaded for the language `#1'. Using the pattern for}%
\typeout{** the default language instead.}%
\else
\language=\csname l@#1\endcsname
\fi
#2}}
\providecommand{\BIBdecl}{\relax}
\BIBdecl

\bibitem{holtryd2022sok}
N.~R. Holtryd, M.~Manivannan, and P.~Stenstr{\"o}m, ``Sok: Analysis of root causes and defense strategies for attacks on microarchitectural optimizations,'' \emph{arXiv preprint arXiv:2212.10221}, 2022.

\bibitem{Intel}
Intel, ``{Analysis of Speculative Execution Side Channels},'' \url {https://software.intel.com/sites/default/files/managed/b9/f9/336983-Intel-Analysis-of-Speculative-Execution-Side-Channels-White-Paper.pdf}, 2018.

\bibitem{liu2015last}
F.~Liu, Y.~Yarom, Q.~Ge, G.~Heiser, and R.~B. Lee, ``Last-level cache side-channel attacks are practical,'' in \emph{{IEEE Symposium on Security and Privacy}}, 2015.

\bibitem{Lipp2018}
M.~Lipp, M.~Schwarz, D.~Gruss, T.~Prescher, W.~Haas, A.~Fogh, J.~Horn, S.~Mangard, P.~Kocher, D.~Genkin, Y.~Yarom, and M.~Hamburg, ``{Meltdown: Reading Kernel Memory from User Space},'' in \emph{Usenix Security}, 2018.

\bibitem{Bulck2018}
J.~V. Bulck, M.~Minkin, O.~Weisse, D.~Genkin, B.~Kasikci, F.~Piessens, M.~Silberstein, T.~Wenisch, Y.~Yarom, and R.~Strackx, ``{Foreshadow: Extracting the Keys to the Intel SGX Kingdom with Transient Out-of-Order Execution},'' in \emph{Usenix Security}, 2018.

\bibitem{Kocher2018}
P.~Kocher, J.~Horn, A.~Fogh, D.~Genkin, D.~Gruss, W.~Haas, M.~Hamburg, M.~Lipp, S.~Mangard, T.~Prescher, M.~Schwarz, and Y.~Yarom, ``{Spectre Attacks: Exploiting Speculative Execution},'' in \emph{S\&P}, 2019.

\bibitem{Koruyeh2018}
E.~M. Koruyeh, K.~N. Khasawneh, C.~Song, and N.~Abu-Ghazaleh, ``{Spectre Returns! Speculation Attacks using the Return Stack Buffer},'' in \emph{WOOT}, 2018.

\bibitem{RIDL}
S.~van Schaik, A.~Milburn, S.~Österlund, P.~Frigo, G.~Maisuradze, K.~Razavi, H.~Bos, and C.~Giuffrida, ``{RIDL: Rogue In-flight Data Load},'' in \emph{S\&P}.\hskip 1em plus 0.5em minus 0.4em\relax {IEEE}, 2019.

\bibitem{Fallout}
C.~Canella, D.~Genkin, L.~Giner, D.~Gruss, M.~Lipp, M.~Minkin, D.~Moghimi, F.~Piessens, M.~Schwarz, B.~Sunar, J.~Van~Bulck, and Y.~Yarom, ``{Fallout}: Leaking data on {Meltdown}-resistant {CPUs},'' in \emph{CCS}, 2019.

\bibitem{ZombieLoad}
M.~Schwarz, M.~Lipp, D.~Moghimi, J.~Van~Bulck, J.~Stecklina, T.~Prescher, and D.~Gruss, ``{ZombieLoad} : Cross-privilege-boundary data sampling,'' in \emph{CCS}, 2019.

\bibitem{plundervolt}
K.~Murdock, D.~Oswald, F.~D. Garcia, J.~Van~Bulck, D.~Gruss, and F.~Piessens, ``Plundervolt: Software-based fault injection attacks against intel sgx,'' in \emph{2020 IEEE Symposium on Security and Privacy (SP)}.\hskip 1em plus 0.5em minus 0.4em\relax IEEE, 2020, pp. 1466--1482.

\bibitem{clkscrew}
A.~Tang, S.~Sethumadhavan, and S.~Stolfo, ``$\{$CLKSCREW$\}$: Exposing the perils of $\{$Security-Oblivious$\}$ energy management,'' in \emph{26th USENIX Security Symposium (USENIX Security 17)}, 2017, pp. 1057--1074.

\bibitem{qiu2019voltjockey}
P.~Qiu, D.~Wang, Y.~Lyu, and G.~Qu, ``Voltjockey: Breaking sgx by software-controlled voltage-induced hardware faults,'' in \emph{2019 Asian Hardware Oriented Security and Trust Symposium (AsianHOST)}.\hskip 1em plus 0.5em minus 0.4em\relax IEEE, 2019, pp. 1--6.

\bibitem{voltjockey-TZ}
------, ``Voltjockey: Breaching trustzone by software-controlled voltage manipulation over multi-core frequencies,'' in \emph{Proceedings of the 2019 ACM SIGSAC Conference on Computer and Communications Security}, 2019, pp. 195--209.

\bibitem{voltjockey-SGX}
P.~Qiu, D.~Wang, Y.~Lyu, R.~Tian, C.~Wang, and G.~Qu, ``Voltjockey: A new dynamic voltage scaling-based fault injection attack on intel sgx,'' \emph{IEEE Transactions on Computer-Aided Design of Integrated Circuits and Systems}, vol.~40, no.~6, pp. 1130--1143, 2020.

\bibitem{voltpwn}
Z.~Kenjar, T.~Frassetto, D.~Gens, M.~Franz, and A.-R. Sadeghi, ``$\{$V0LTpwn$\}$: Attacking x86 processor integrity from software,'' in \emph{29th USENIX Security Symposium (USENIX Security 20)}, 2020, pp. 1445--1461.

\bibitem{bernsteincache}
D.~J. Bernstein, ``Cache-timing attacks on {AES},'' 2005.

\bibitem{Osvik2006}
D.~A. Osvik, A.~Shamir, and E.~Tromer, ``{Cache Attacks and Countermeasures: The Case of AES},'' in \emph{CT-RSA}, 2006.

\bibitem{Yarom14}
Y.~Yarom and K.~Falkner, ``{Flush+Reload: A High Resolution, Low Noise, L3 Cache Side-channel Attack},'' in \emph{Usenix Security}, 2014.

\bibitem{gruss2016flush+}
D.~Gruss, C.~Maurice, K.~Wagner, and S.~Mangard, ``{Flush+ Flush}: a fast and stealthy cache attack,'' in \emph{International Conference on Detection of Intrusions and Malware, and Vulnerability Assessment}.\hskip 1em plus 0.5em minus 0.4em\relax Springer, 2016, pp. 279--299.

\bibitem{briongos2020reload+}
S.~Briongos, P.~Malag{\'o}n, J.~M. Moya, and T.~Eisenbarth, ``{RELOAD+ REFRESH}: Abusing cache replacement policies to perform stealthy cache attacks,'' in \emph{29th USENIX Security Symposium (USENIX Security 20)}, 2020, pp. 1967--1984.

\bibitem{saileshwar2021streamline}
G.~Saileshwar, C.~W. Fletcher, and M.~Qureshi, ``Streamline: a fast, flushless cache covert-channel attack by enabling asynchronous collusion,'' in \emph{Proceedings of the 26th ACM International Conference on Architectural Support for Programming Languages and Operating Systems}, 2021, pp. 1077--1090.

\bibitem{cui2022abusing}
Y.~Cui and X.~Cheng, ``Abusing cache line dirty states to leak information in commercial processors,'' in \emph{2022 IEEE International Symposium on High Performance Computer Architecture (HPCA)}.\hskip 1em plus 0.5em minus 0.4em\relax IEEE, 2022.

\bibitem{guo2022adversarial}
Y.~Guo, A.~Zigerelli, Y.~Zhang, and J.~Yang, ``Adversarial prefetch: New cross-core cache side channel attacks,'' in \emph{2022 IEEE Symposium on Security and Privacy (SP)}.\hskip 1em plus 0.5em minus 0.4em\relax IEEE, 2022, pp. 1458--1473.

\bibitem{aldaya2019port_portsmash}
A.~C. Aldaya, B.~B. Brumley, S.~ul~Hassan, C.~P. Garc{\'\i}a, and N.~Tuveri, ``Port contention for fun and profit,'' in \emph{2019 IEEE Symposium on Security and Privacy (SP)}.\hskip 1em plus 0.5em minus 0.4em\relax IEEE, 2019, pp. 870--887.

\bibitem{lipp2021platypus}
M.~Lipp, A.~Kogler, D.~Oswald, M.~Schwarz, C.~Easdon, C.~Canella, and D.~Gruss, ``Platypus: Software-based power side-channel attacks on x86,'' in \emph{2021 IEEE Symposium on Security and Privacy (SP)}.\hskip 1em plus 0.5em minus 0.4em\relax IEEE, 2021, pp. 355--371.

\bibitem{qiu2022pmu}
P.~Qiu, Q.~Gao, D.~Wang, Y.~Lyu, C.~Liu, X.~Li, C.~Wang, and G.~Qu, ``Pmu-spill: Performance monitor unit counters leak secrets in transient executions,'' in \emph{2022 Asian Hardware Oriented Security and Trust Symposium (AsianHOST)}.\hskip 1em plus 0.5em minus 0.4em\relax IEEE, 2022, pp. 1--6.

\bibitem{canella2019sokattacks}
C.~Canella, J.~Van~Bulck, M.~Schwarz, M.~Lipp, B.~Von~Berg, P.~Ortner, F.~Piessens, D.~Evtyushkin, and D.~Gruss, ``A systematic evaluation of transient execution attacks and defenses,'' in \emph{28th USENIX Security Symposium (USENIX Security 19)}, 2019.

\bibitem{abel2019uops}
A.~Abel and J.~Reineke, ``{uops.info: Characterizing latency, throughput, and port usage of instructions on Intel microarchitectures},'' in \emph{ASPLOS}, 2019.

\bibitem{intelsys}
{Intel Corporation}, \emph{{Intel\textsuperscript {\textregistered} 64 and IA-32 Architectures Software Developer's Manual}}, 2019.

\bibitem{Degenbaev2012}
U.~Degenbaev, ``{Formal Specification of the x86 Instruction Set Architecture},'' Ph.D. dissertation, Universit{\"a}t des Saarlandes, 2012.

\bibitem{WikiChip}
{WikiChip}, ``{Skylake (client) - Microarchitectures - Intel},'' \url {https://en.wikichip.org/wiki/intel/microarchitectures/skylake_(client)}, 2018, accessed: May, 2021.

\bibitem{Gras2020}
B.~Gras, C.~Giuffrida, M.~Kurth, H.~Bos, and K.~Razavi, ``{ABSynthe: Automatic Blackbox Side-channel Synthesis on Commodity Microarchitectures},'' in \emph{NDSS}, 2020.

\bibitem{Weber2021}
D.~Weber, A.~Ibrahim, H.~Nemati, M.~Schwarz, and C.~Rossow, ``Osiris: Automated discovery of microarchitectural side channels,'' in \emph{Usenix Security}, 2021.

\bibitem{Lustig2014}
D.~Lustig, M.~Pellauer, and M.~Martonosi, ``Pipecheck: Specifying and verifying microarchitectural enforcement of memory consistency models,'' in \emph{MICRO}, 2014.

\bibitem{Rodrigo2021}
J.~R.~S. Vicarte, P.~Shome, N.~Nayak, C.~Trippel, A.~Morrison, D.~Kohlbrenner, and C.~W. Fletcher, ``{Opening Pandora's Box: A Systematic Study of New Ways Microarchitecture Can Leak Private Data},'' in \emph{ISCA}, 2021.

\bibitem{gras2018translation}
B.~Gras, K.~Razavi, H.~Bos, and C.~Giuffrida, ``Translation leak-aside buffer: Defeating cache side-channel protections with {TLB} attacks,'' in \emph{27th USENIX Security Symposium (USENIX Security 18)}, 2018, pp. 955--972.

\bibitem{branchscope}
D.~Evtyushkin, R.~Riley, N.~C. Abu-Ghazaleh, ECE, and D.~Ponomarev, ``Branchscope: A new side-channel attack on directional branch predictor,'' in \emph{Proceedings of the Twenty-Third International Conference on Architectural Support for Programming Languages and Operating Systems}, ser. ASPLOS '18, 2018, pp. 693--707.

\bibitem{coronado2021branchboozle}
A.~R.~H. Coronado, W.~Lee, and W.-M. Lin, ``Branchboozle: a side-channel within a hidden pattern history table of modern branch prediction units.'' in \emph{SAC}, 2021, pp. 1617--1625.

\bibitem{VanBulck2020}
J.~{Van Bulck}, D.~Moghimi, M.~Schwarz, M.~Lipp, M.~Minkin, D.~Genkin, Y.~Yarom, B.~Sunar, D.~Gruss, F.~Piessens, and K.~Leuven, ``{LVI: Hijacking Transient Execution through Microarchitectural Load Value Injection},'' in \emph{S{\&}P}, 2020.

\bibitem{Ragab2021}
H.~Ragab, A.~Milburn, K.~Razavi, H.~Bos, and C.~Giuffrida, ``{CrossTalk: Speculative Data Leaks Across Cores Are Real},'' in \emph{S\&P}.\hskip 1em plus 0.5em minus 0.4em\relax {IEEE}, 2021.

\bibitem{Maisuradze2018}
G.~Maisuradze and C.~Rossow, ``{ret2spec: Speculative Execution Using Return Stack Buffers},'' in \emph{CCS}, 2018.

\bibitem{Google2018}
P.~Z. Google, ``{Speculative Execution, Variant 4: Speculative Store Bypass},'' \url {https://bugs.chromium.org/p/project-zero/issues/detail?id=1528}, 2018, accessed: May, 2021.

\bibitem{Kiriansky2018}
V.~Kiriansky and C.~Waldspurger, ``{Speculative Buffer Overflows: Attacks and Defenses},'' \emph{arXiv}, 2018.

\bibitem{IntelVulnList}
{Intel}, ``{Affected Processors: Transient Execution Attacks \& Related Security Issues by CPU},'' \url{https://www.intel.com/content/www/us/en/developer/topic-technology/software-security-guidance/processors-affected-consolidated-product-cpu-model.html}, 2022, accessed: August, 2022.

\bibitem{cauligi2021sok}
S.~Cauligi, C.~Disselkoen, D.~Moghimi, G.~Barthe, and D.~Stefan, ``{SoK: Practical Foundations for Spectre Defenses},'' in \emph{S\&P}.\hskip 1em plus 0.5em minus 0.4em\relax {IEEE}, 2022.

\bibitem{kadir2019retpoline}
M.~F.~A. Kadir, J.~K. Wong, F.~Ab~Wahab, A.~F. A.~A. Bharun, M.~A. Mohamed, and A.~H. Zakaria, ``Retpoline technique for mitigating spectre attack,'' in \emph{2019 6th International Conference on Electrical and Electronics Engineering (ICEEE)}.\hskip 1em plus 0.5em minus 0.4em\relax IEEE, 2019, pp. 96--101.

\bibitem{gonzalez2019replicating}
A.~Gonzalez, B.~Korpan, J.~Zhao, E.~Younis, and K.~Asanovi{\'c}, ``{Replicating and Mitigating Spectre Attacks on an Open Source RISC-V Microarchitecture},'' in \emph{CARRV}, 2019.

\bibitem{Li2022}
C.~Li and J.-L. Gaudiot, ``Detecting spectre attacks using hardware performance counters,'' \emph{IEEE Transactions on Computers}, 2022.

\bibitem{Wang2020}
G.~Wang, S.~Chattopadhyay, A.~K. Biswas, T.~Mitra, and A.~Roychoudhury, ``{KLEESpectre: Detecting information leakage through speculative cache attacks via symbolic execution},'' \emph{TOSEM}, 2020.

\bibitem{Guarnieri2018}
M.~Guarnieri, B.~K{\"o}pf, J.~Morales, J.~Reineke, and A.~Sanchez, ``{Spectector: Principled Detection of Speculative Information Flows},'' in \emph{S\&P}.\hskip 1em plus 0.5em minus 0.4em\relax {IEEE}, 2020.

\end{thebibliography}

\section*{Appendix}

\subsection*{Appendix A: Victim}

The victim code is shown in Fig.~\ref{fig:victim}, as same as the POC in Meltdown\cite{Lipp2018}. We run it parallelly with the attacker in the same physical core but a different logical core for a higher reading rate. The victim will try to keep the secret string cached in the cache line.

\begin{figure}[ht]
\begin{center}
    \begin{minipage}{0.92\columnwidth}
    \begin{lstlisting}[style=short_code, language=c]
char *strings[] = {SECRET_STR};

while (1)
{
    // keep string cached for better results
    volatile size_t dummy = 0, i;
    for (i = 0; i < len; i++)
    {
        dummy += secret[i];
    }
    sched_yield();
}\end{lstlisting}
    \captionof{lstlisting}{Code Snippet for victim.}
    \label{fig:victim}
    \end{minipage}
\end{center}
\end{figure}

\end{document}